\documentclass[11pt,aps,preprint,showkeys,showpacs,amsmath,amssymb,amsfonts,eqnarray,array]{revtex4}
\usepackage{anysize}
\marginsize{1.9cm}{1.9cm}{1.55cm}{1.55cm}
\usepackage{graphicx}
\usepackage{dcolumn}
\usepackage{bm}
\begin{document}


\title{The Quantum Pinch Effect in Semiconducting Quantum Wires: A Bird's-Eye View \\}
\author{Manvir S. Kushwaha}
\affiliation
{\centerline {Department of Physics and Astronomy, Rice University, P.O. Box 1892, Houston, TX 77251, USA}}


\date{\today}
\begin{abstract}

Those who measure success with culmination do not seem to be aware that life is a journey not a
destination. This spirit is best reflected in the unceasing failures in efforts for solving the
problem of controlled thermonuclear fusion for even the simplest pinches for over decades; and
the nature keeps us challenging with examples. However, these efforts have permitted researchers
the obtention of a dense plasma with a lifetime that, albeit short, is sufficient to study the
physics of the pinch effect, to create methods of plasma diagnostics, and to develop a modern
theory of plasma processes. Most importantly, they have impregnated the solid state plasmas,
particularly the electron-hole plasmas in semiconductors, which do not suffer from the issues
related with the confinement and which have demonstrated their potential not only for the
fundamental physics but also for the device physics. Here, we report on a two-component,
cylindrical, quasi-one-dimensional quantum plasma subjected to a {\em radial} confining harmonic
potential and an applied magnetic field in the symmetric gauge. It is demonstrated that such a
system as can be realized in semiconducting quantum wires offers an excellent medium for observing
the quantum pinch effect at low temperatures. An exact analytical solution of the problem allows
us to make significant observations: surprisingly, in contrast to the classical pinch effect, the
particle density as well as the current density display a {\em determinable} maximum before
attaining a minimum at the surface of the quantum wire. The effect will persist as long as the
equilibrium pair density is sustained. Therefore, the technological promise that emerges is the
route to the precise electronic devices that will control the particle beams at the nanoscale.

\end{abstract}

\maketitle

\section{A Kind of Introduction}

The greatest scientific challenge in the history of civilization is known to be set forth with ancient belief
in the five {\em basic} elements: earth, water, air, fire, and aether [see, e.g., the Yoonir in Fig. 1]. Early
credence in these basic elements dates back to pre-Socratic times and persisted throughout the Middle Ages and
into the Renaissance, deeply influencing European thought and culture. These five elements are also sometimes
associated with the five platonic solids. Many philosophies have a set of five elements believed to reflect
the simplest essential principles upon which the constitution of everything is based. Most frequently, five
elements refer to the ancient concepts which some science writers compare to the present-day states of matter,
relating earth to solid, water to liquid, air to gas, and fire to {\em plasma}. Historians trace the evolution
of modern theory of chemical elements, as well as chemical compounds, to medieval and Greek models. According
to Hinduism, the five elements are found in {\em Vedas}, especially {\em Ayurveda}; are associated with the
five senses: (i) hearing, (ii) touch, (iii) sight, (iv) taste, and (v) smell; and act as the gross medium for
the experience of sensations. They further suggest that all of creation, including the human body, is made up
of these five elements and that, upon death, the human body dissolves into these five elements, thereby
balancing the divine cycle of nature.

\begin{figure}[htbp]
\includegraphics*[width=8cm,height=8cm]{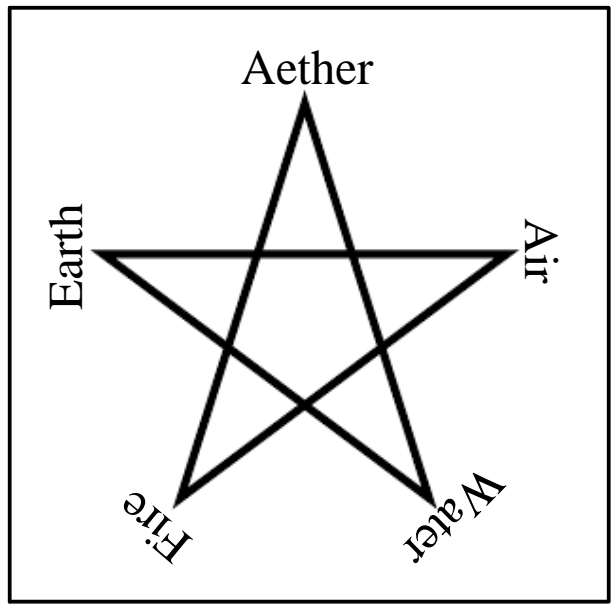}
\centering
\caption{(Color online) The Yoonir [or, more formally, the Star of Yoonir] is one of the most important and
sacred cosmological symbols in the Serer religion. It is the brightest star of the universe and commonly
known as the Star of Sirius. Because of its uniqueness, I choose it to represent the five {\em basic}
elements of life.}
\label{fig1}
\end{figure}

This review article has very much to do with the plasma -- another ghostly gift of Mother Nature -- a state of
matter whose size and/or shape is beyond imagination. A plasma is, by definition, an ionized gas in which the
length which separates the single-particle behavior from the collective behavior is smaller than the
characteristic lengths of interest. Researchers in plasma physics have come to the recognition that $99.999\%$
of the mass of the universe exists in the form of plasma. The term plasma refers to the {\em blood} of the
universe. Plasma is invisible; pervades all space in gigantic streams; powers everything: every galaxy, every
star; formed by the nature of its dynamics; and its organization in space is shaped by the electromagnetic
force of the universe. The same force also pinches the streams ever tighter, until the concentration becomes
so great that {\em explosive} node-points result where plasma dissipates. Our Sun is known to be located at the
center of one of the node points [see Fig. 2]. Physically, plasma is loosely defined as globally neutral medium
made up of unbound electrically charged particles. The term unbound, by no means, refers to the particles being
free from experiencing forces. When the charges move, they generate electric currents which create magnetic
fields which, in turn, act upon the currents and, as a result, they are affected by each other’s fields. Much of
the understanding of plasmas has come from the pursuit of controlled nuclear fusion and fusion power, for which
plasma physics provides the scientific basis. As to the interpretation of (space) plasma dynamics, the
magnetohydrodynamics (MHD) that provides theoretical framework has always stood to the test.

\begin{figure}[htbp]
\includegraphics*[width=8cm,height=8cm]{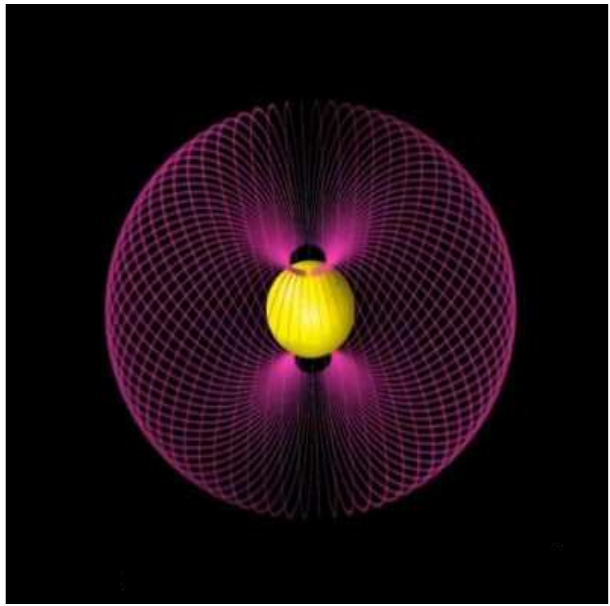}
\centering
\caption{(Color online) The geometry of the Sun at a plasma node. At this point on, plasma expands again and
moves on, collecting more plasma as it flows around in the space. Instead of plasma exploding at the node
point of the Sun, the Sun consumes the plasma... [After the Home page of Rolf A. Witzsche.]}
\label{fig2}
\end{figure}

Literature is a live witness that pinches were the first device used by mankind for controlled nuclear fusion.
{\em Pinch effect is the manifestation of self-constriction of plasma caused by the action of a magnetic field
that is generated by the parallel electric currents}. The result of this manifestation is that the plasma is
heated as well as confined. The phenomenon is also referred to as Bennett pinch, electromagnetic pinch, plasma
pinch, magnetic pinch, or (most commonly) the pinch effect. Pinch effect takes place naturally in electrical
discharges such as lightning bolts [see, e.g., Fig. 3], the auroras, current sheets, and solar flares. Pinches
may become unstable. They radiate energy as light across the whole electromagnetic spectrum including x-rays,
radio waves, gamma rays, synchrotron radiation, and visible light. They also produce neutrons, as a result of
fusion -- although a desired nuclear fusion has never been achieved because the confinement and the instability
are two concurrent and diametric events. As to the applications, pinches are used to generate x-rays and the
intense magnetic fields generated are used in forming of the metals [see, e.g., Fig. 4]. They also
have applications in astrophysical studies, space propulsion, and particle beam weapons. Numerous large pinch
machines have been built in order to study the fusion power. Clearly, a cylindrical geometry is pertinent to
the effect.

\begin{figure}[htbp]
\includegraphics*[width=8cm,height=8cm]{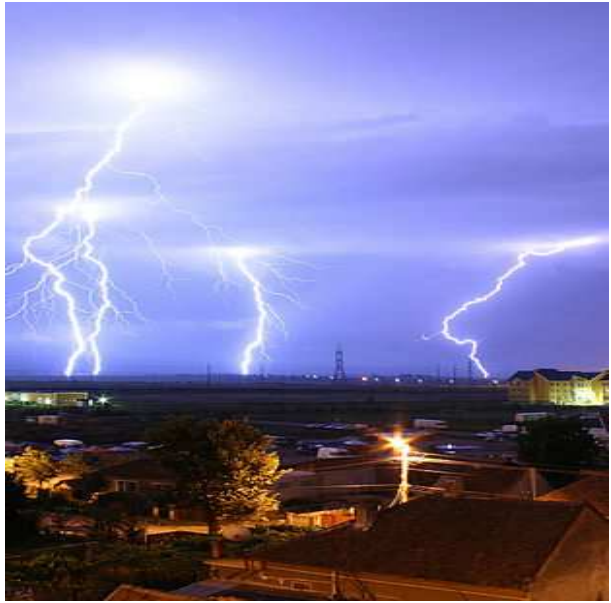}
\centering
\caption{(Color online) Lightning bolts illustrating electromagnetically pinched plasma filaments is an example
of plasma present at Earth's surface. Typically, lightning discharges 30,000 amperes at up to 100 million volts,
and emits light, radio waves, X-rays and even gamma rays. Plasma temperatures in lightning can approach 28,000 K
(28,000 C; 50,000 F) and electron densities may exceed $10^{24}$ m$^{-3}$. [Courtesy of the Wikipedia].}
\label{fig3}
\end{figure}


The magnetic self-constriction (or self-attraction) of the parallel electric currents is the best-known and the
simplest example of Amp\`ere's force law, which states that the force per unit length between two straight,
parallel, conducting wires is given by
\begin{align}
{\frac {F_{m}}{L}}=2k_{A}{\frac {I_{1}I_{2}}{r}}\, ,
\end{align}
where $k_A$ is the magnetic force constant from the Biot-Savart law, $F_m/L$ is the total force on either wire
per unit length (of the shorter), $r$ is the distance between the two wires, and $I_1$, $I_2$ are the direct
currents carried by the wires. This is a good approximation if one wire is sufficiently longer than the other
that it can be approximated as infinitely long, and if the distance between the wires is small compared to their
lengths (so that the one infinite-wire approximation holds), but large compared to their diameters (so that they
may also be approximated as infinitely thin lines). The value of $k_A$ depends upon the system of units chosen
and decides how large the unit of current will be. In the SI system, $k_{A} = \mu _{0}/{4\pi}$, with
$\mu _{0} = 4\pi \times 10^{-7}$ N/A$^2$; whereas in the CGS system, $k_A=1$. For the modest current levels of
a few amperes, this force is usually negligible, but when current levels approach a million amperes, such as in electrochemistry, this force can be significant since the pressure is proportional to $I^2$. The form of
Amp\`ere's force law commonly given was derived by Maxwell and is one of the several expressions consistent with
the original experiments of Amp\`ere and Gauss. Note in passing that the Amp\`ere's law underlies the definition
of the ampere, the SI unit of current.

\begin{figure}[htbp]
\includegraphics*[width=6cm,height=8cm]{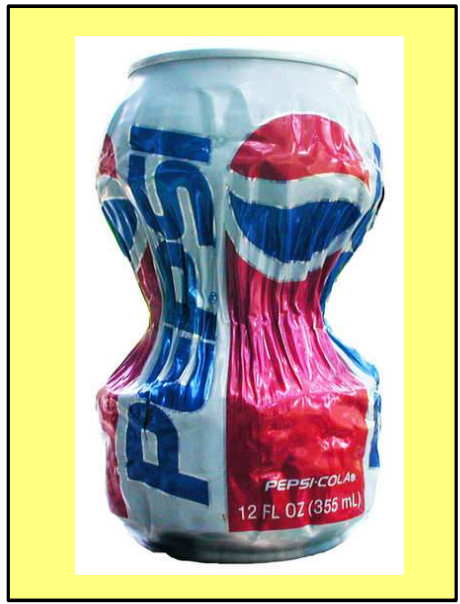}
\centering
\caption{(Color online) Electromagnetically pinched aluminium can, produced from a pulsed magnetic field
created by rapidly discharging 2 kilojoules from a high voltage capacitor bank into a 3-turn coil of heavy
gauge wire. [After Bert Hickman, Stoneridge Engineering].}
\label{fig4}
\end{figure}

Let us scrutinize how and why the pinch effect comes into being. For this purpose, we consider a cylindrical
coordinate frame with the axis z along the axis of the wire. For the moment, we do not even need to know
its size such as the radius. We assume that the electrons (of charge -$e$, with $e>0$) flow in the positive
direction of the z axis with a velocity ${\bf v}>0$. For the symmetry reasons, we restrict the magnetic
field to be azimuthal, i.e., to have only component $B_{\theta}=B_{\theta}(r)$. In order to evaluate
$B_{\theta}$, we use Ampere's law in the integral form for the circular path as depicted in Fig. 5. The
result is that $B_{\theta}=(2\pi r/c)\,J$, where $J=-n_e e v < 0$. Therefore, $B_{\theta} < 0$, i.e., it
is oriented clockwise with respect to the z axis. The magnetic force
${\bf F}=-(e/c)\,({\bf v}\times{\bf B})=-2\pi n_e e^2 (v^2/c^2) {\bf r}$ is in the radial direction guided
towards the axis. The dependence on $e^2$ implies that the force is independent of the sign of the charge.
Clearly then, the magnetic force tends to push the electrons towards the axis of the wire. In other words,
in a flowing current beam the self-generated magnetic field always acts to {\em pinch} the beam -- or to
let it shrink around the axis. Next, the Lorentz force
${\bf F}=-e\,({\bf E}+(1/c)({\bf v}\times{\bf B}))$ yields ${\bf E}=-(1/c)\,({\bf v}\times{\bf B})$ for
${\bf F}$ to be zero. This yields ${\bf E}=-2\pi n_e e (v^2/c^2)\,{\bf r}$. From the integral form of the
Gauss theorem, we find that this field is generated by a charge density $\rho$, which is uniform over the
wire such that $E=2\pi r \rho$, which gives $\rho=-n_e e (v^2/c^2)$. Since $\rho=e\,(n_i-n_e)$ holds, we
obtain $n_e= n_i /[1-v^2/c^2]$; $n_i$ being the ion density. The electron density is uniform, but it
exceeds the value $n_e=n_i$ that would ensure $\rho = 0$. This leads us to infer that there has been some
accumulation of electrons inside the wire, rising the density by a factor $(1-v^2/c^2)$. For an electron
in an Ohmic conductor usually $v\approx 1$ cm/sec, thus $v^2/c^2\approx 10^{20}$ implying that the
pinch effect is too small to be observed. The effect can be robust in high density particle beam or plasma
columns where $v/c$ is non-negligible.

\begin{figure}[htbp]
\includegraphics*[width=8cm,height=5cm]{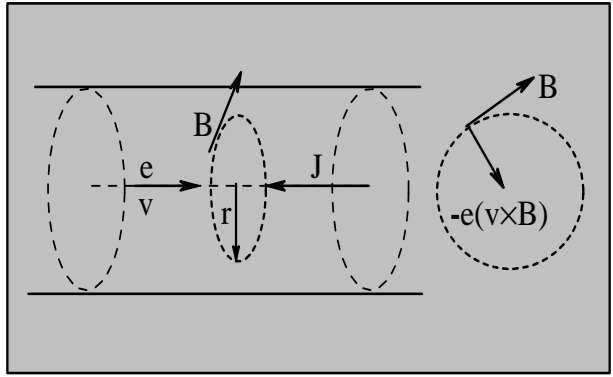}
\centering
\caption{(Color online) A cylindrical geometry utilized to demonstrate how and why the pinch effect is
[and can be] a real deal given the appropriate plasma medium where $v^2/c^2$ is not much less than one.}
\label{fig5}
\end{figure}

Characteristically, the pinch effect can be understood most easily through the example of a current flowing
along the axis of a cylinder filled with a conducting medium. The lines of force of the magnetic field
${\bf B}$ generated by the current have the form of the concentric circles whose planes are perpendicular
to the axis of the cylinder. The electrodynamic force per unit volume of a conducting medium with a current
density ${\bf J}$ is ${\bf F}=(1/c)$(${\bf J} \times {\bf B}$); this force is guided towards the axis of the
cylinder and tends to squeeze down the medium. The state that arises is the pinch effect. The pinch effect
may also be considered a simple consequence of Amp\ ere's law [see above], which describes the magnetic
attraction between the individual parallel current filaments whose aggregate is the current cylinder. Notice
that the magnetic pressure ($P_{m}$) is impeded by the kinetic pressure ($P_{k}$) -- directed from the
axis -- of the medium arising from the thermal motion of the medium's particles. When the current is
sufficiently large such that the {\em plasma} $\beta$ [$=P_k/P_m$] $<1$, the pinch effect comes into being.
The pinch effect requires that the population of the charge carriers of opposite signs in the medium be
nearly equal. In a medium of charge carriers of single sign, the electric field of the space charge
effectively impedes the inward action of the magnetic pressure. It is obvious by now that the pinch effect is
paramount in the problems of controlled thermonuclear fusion. As to the geometries [see Appendix A], the term $\theta$-pinch has come into wide usage to denote an important confinement scheme which relies on the
repulsion of oppositely directed currents and which is thus not in accord with the original definition of the
pinch effect -- self-attraction of the parallel currents. As such, the z-pinch remains by far the most
preferred scheme for the plasma confinement [see Fig. 6].

\begin{figure}[htbp]
\includegraphics*[width=8cm,height=5cm]{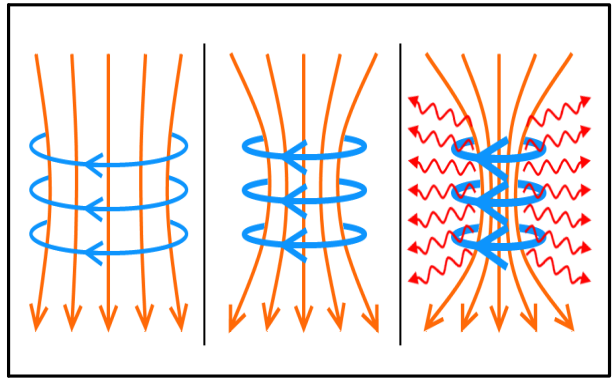}
\centering
\caption{(Color online) Schematics of the Z-pinch: A current (orange) generates a magnetic field (blue),
which causes the current to pinch inwards along the axis by way of the Biot-Savart force:
${\bf F}=(1/c)$(${\bf J} \times {\bf B}$). This amplifies the magnetic field and accelerates the pinch,
heating the plasma and causing it to radiate X-rays (red).}
\label{fig6}
\end{figure}

In order to take this brief history of the pinch effect to the point where it was formally discussed first,
we would like to recall the mechanism behind the Bennett pinch in the gaseous plasmas. Incidentally, the
Bennett pinch is also known by other names such as equilibrium or steady-state pinch because in the Bennett
pinch no variable is a function of time. As stated above, while the magnetic pressure tends to confine the
plasma, the kinetic pressure tends to oppose it. When these forces are balanced, an equilibrium is attained.
The resulting state is what we call the Bennett pinch. Therefore, the use of the steady-state MHD model in
which $\nabla p =(1/c)$ (${\bf J} \times {\bf B}$) makes sense. Here $p=n k_B$($T_e + T_i$) is the kinetic
pressure of the plasma particles due to its thermal motion -- as a reaction to the magnetic force. Symbols
$n=$ ($n_e=n_i$), $k_B$, and $T_e$ ($T_i$) are, respectively, the particle density, Boltzamann's constant,
and electron (ion) temperature. Notice that $p$, ${\bf J}$, and ${\bf B}$ are all local ($r$-dependent)
functions. The equilibrium z-pinch can be described by the balance of kinetic pressure at the axis and
magnetic pressure at the surface ($r=a$):
\begin{align}
n\, k_B\, (T_e + T_i)=-\frac{B^2_{\theta}(a)}{8\,\pi}\, .
\end{align}
The magnetic field at the cylindrical surface is calculated from the total current by using the Amp\`ere's
law to write
\begin{align}
B_{\theta}(a)= \frac{2\,I}{c\,a}\, .
\end{align}
These two equations allow us to determine the relationship between the pinch radius, the temperature at the
axis, and the total current:
\begin{align}
(T_e + T_i) \propto \frac{I^2}{a^2}\, .
\end{align}
This is known as the Bennett-relation after Willard H. Bennett who first discussed this z-pinch effect in
1934. Equation (4) estimates the total current that must be discharged through the plasma medium in order
to confine the plasma at a specified temperature for a given linear particle density and the pinch radius
$a$. Figure 7 illustrates his typical chart showing the total current (I) versus the number of particles
per unit length (N) in four distinct regimes. The conducting medium is plasma which is assumed to be
non-rotational and $P_k$ at the surface is much smaller than near the axis.

\begin{figure}[htbp]
\includegraphics*[width=8cm,height=8cm]{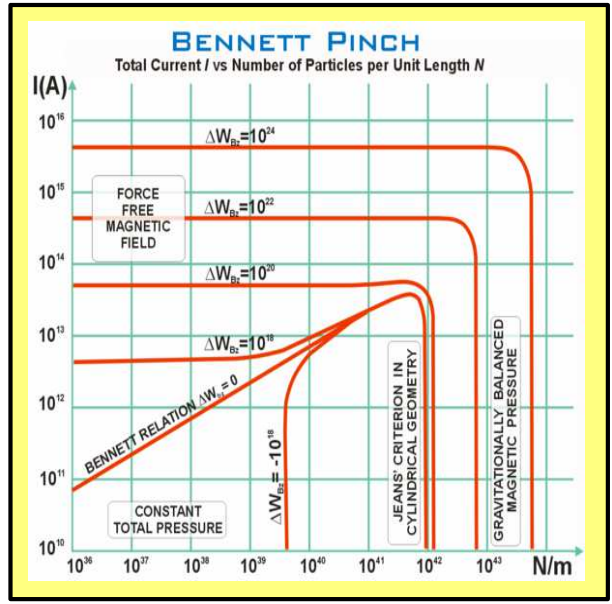}
\centering
\caption{(Color online) The static Z-pinch showing the total current (I) versus the number of particles per
unit length (N) -- first studied by Bennett in 1934. The chart illustrates four physically distinct regions.
The plasma temperature is 20 K, the mean particle mass $3\times 10^{-27}$ kg, and $\Delta W B_z$ is the
excess magnetic energy per unit length due to the axial magnetic field $B_z$. The plasma is assumed to be
non-rotational, and the kinetic pressure at the edges is much smaller than inside. [Courtesy of the Wikipedia].}
\label{fig7}
\end{figure}

\newpage

\section{The Quantum Pinch Effect in Quantum Wires}
The pinch effect is one of the most fascinating phenomena in plasma physics with immense applications
to the problems of peace and war. It is the manifestation of radial constriction of a compressible
conducting plasma [or a beam of charged particles] due to the magnetic field generated by the parallel
electric currents. Cylindrical symmetry is central to the realization of the effect. In the literature,
the phenomenon is also referred to as self-focusing, magnetic pinch, plasma pinch, or Bennett pinch.
The pinch effect in the cylindrical geometry is classified after the direction in which the current
flows: In a $\theta$-pinch, the current is azimuthal and the magnetic field axial; in a $z$-pinch, the
current is axial and the magnetic field azimuthal; in a screw (or $\theta$-$z$) pinch, an effort is made
to combine both $\theta$-pinch and $z$-pinch [see Fig. 8 and Appendix A].

\begin{figure}[htbp]
\includegraphics*[width=8cm,height=7.5cm]{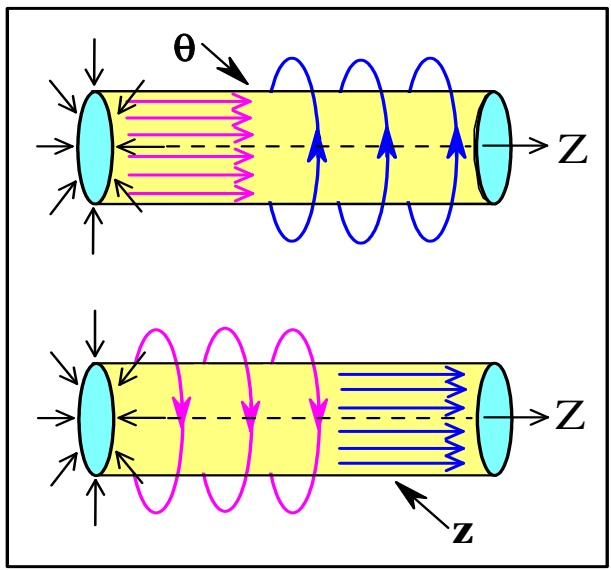}
\centering
\caption{(Color online) The schematics of the $\theta$-pinch (the upper panel) and the $z$-pinch
(the lower panel). The curves in blue (magenta) refer to the current (magnetic field).
[After Kushwaha, Ref. 39].}
\label{fig8}
\end{figure}

While a popular reference dates back to 1790 when Martinus van Marum in Holland created an explosion by
discharging 100 Leyden jars into a wire, the first formal study of the effect was not undertaken until
1905 when Pollock and Barraclough [1] investigated a compressed and distorted copper tube after it had
been struck by lightning [see Fig. 9]. They argued that the magnetic forces due to the large current
flow could have caused the compression and distortion. Shortly thereafter, Northrupp published a similar
but independent diagnosis of the phenomenon in liquid metals [2]. However, the major breakthrough in the
topic came with the publication of the derivation of the radial pressure balance in a static z-pinch by
Bennett [3]. Curiously enough, the term ``pinch effect" was only coined in 1937 by Tonks in his work on
the high current densities in low pressure arcs [4]. This was followed by a \emph{first} patent for a
fusion reactor based on a toroidal z-pinch submitted by Thompson and Blackman [5]. The subsequent
progress -- theoretical and experimental -- on the pinch effect in the gas discharges was driven by the
quest for the controlled nuclear fusion.

Needless to say, the plasma commands a glamorous status in the physics literature by representing the
{\em fourth} state of matter. Our key interest here is in the solid state plasmas (SSP), which share
several characteristic features with but are also known to have quantitative differences from the
gaseous space plasmas (GSP). The GSP is ill-famed for two inherently juxtaposing characteristics:
confinement and instability. The SSP is, on the contrary, a uniform, equilibrium system tightly bound
to the lattice where the boundary conditions make no sense, whereas in GSP they can be crucial. Notice
that only two-component, semiconductor SSP with approximately equal number of electrons and holes
can support a significant pinching. For a single mobile carrier the space-charge electric field due to
any alteration in the charge-density will impede such inward motion. Thus a metal with only electrons
contributing to the conduction is a poor option. The 1960's had seen considerable experimental [6-15]
and theoretical [16-19] efforts focused on studying pinch effect in the bulk SSP, with instabilities
manifesting themselves as voltage and current oscillations at frequencies up to 50 MHz.

\begin{figure}[htbp]
\includegraphics*[width=7cm,height=8cm]{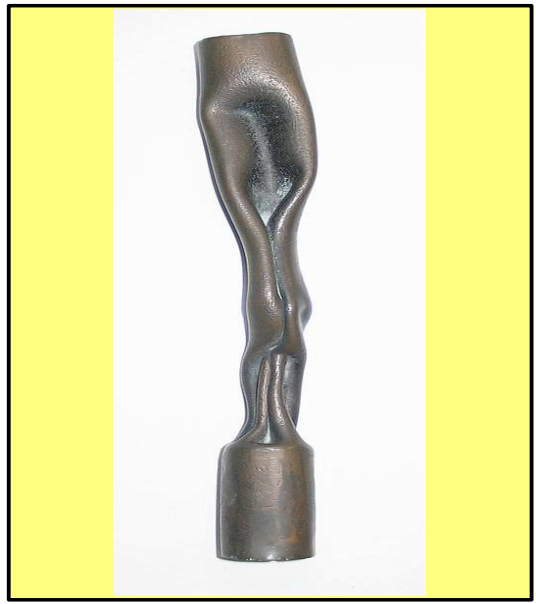}
\centering
\caption{(Color online) A photograph of a section of the crushed copper tube from a lightning rod first
described by Pollock and Barraclough (1905). They correctly described the crushing mechanism as a result
of the interaction of the large current flowing in the conductor with its own magnetic field. [Courtesy
of the Wikipedia].}
\label{fig9}
\end{figure}


Notwithstanding the SSP had been imbued to a greater extent by the research pursuit for the pinch effect
in the GSP, the subject did not gain sufficient momentum because the early 1970's had begun to offer the
condensed matter physicists with the new venues to explore: the semiconducting quantum structures with
reduced dimensions such as quantum wells, quantum wires, quantum dots, and their periodic counterparts.
The continued tremendous research interest in these quasi-n-dimensional electron gas [Q-nDEG] systems can
now safely be accredited to the discovery of the quantum Hall effects -- both integral [20] and fractional
[21]. The fundamental issue behind the excitement in these man-made nanostructures lies in the fact that
the charge carriers exposed to external probes such as electric and/or magnetic fields can exhibit
unprecedented quantal effects that strongly modify their behavior characteristics [22-38]. To what extent
these effects can influence the behavior of a quantum wire [or, more realistically, Q-1DEG system] in the
cylindrical symmetry, which offers a quantum analogue of the classical structure subjected to investigate
the pinch effect in the conventional SSP [or GSP], has not yet been explored. This was the motivation
behind the original work on the quantum pinch effect investigated in semiconducting quantum wires in
Ref. 39.

We consider a two-component, quasi-1DEG system characterized by a radial harmonic confining potential and
subjected to an applied (azimuthal and axial) magnetic field in the cylindrical symmetry. The two-component
Q-1DEG systems are comprised of both types of charge carriers [i.e., electrons and holes] and are known to
have been fabricated in a wide variety of semiconducting quantum wires by optical pumping techniques [40-42].
In the linear response theory, the resultant system is characterized by the single-particle Hamiltonian
$H=H_0+H_1$, where
[after Peierls substitution ${\boldsymbol p}\rightarrow ({\boldsymbol p}\pm\frac{e}{c}{\boldsymbol A})$,
with ${\boldsymbol A}={\boldsymbol A}_0 +{\boldsymbol A}_1$]
\begin{align}
H_0 =& \frac{1}{2}m_{e}{\boldsymbol v}^2_{e} + \frac{1}{2}m_{h}{\boldsymbol v}^2_{h} +
       \frac{1}{2}m_{e}\omega^2_{e} r^2+\frac{1}{2}m_{h}\omega^2_{h}r^2\, , \\
H_1 =& \frac{e}{2 c}[({\boldsymbol v}_{e}\!\cdot\!{\boldsymbol A_1} +
                      {\boldsymbol A_1}\!\cdot\!{\boldsymbol v}_{e})
              \! -\! ({\boldsymbol v}_{h}\!\cdot\!{\boldsymbol A_1} +
                      {\boldsymbol A_1}\!\cdot\!{\boldsymbol v}_{h})]\, ,
\end{align}
to first order in ${\boldsymbol A}_1$, where ${\boldsymbol A}_0$ (${\boldsymbol A}_1$) is the {\em dc} ({\em ac})
part of the vector potential ${\boldsymbol A}$. Here $c$ is the speed of light in vacuum and $-e (+e)$,
$m_e (m_h)$, and $\omega_e (\omega_h)$ are, respectively, the electron (hole) charge, mass, and the characteristic
frequency of the harmonic oscillator.
${\boldsymbol v}_i=\frac{1}{m_i}\,({\boldsymbol p}\pm \frac{e}{c}\,{\boldsymbol A}_0)$ is the velocity operator
for an electron (hole). A few significant aspects of
the formalism are:
(i) the formal Peierls substitution, by which a magnetic field is introduced into the Hamiltonian, is a direct
consequence of gauge invariance under the transformation $\psi \rightarrow \psi e^{i\phi}$, with $\phi$ being
an arbitrary phase,
(ii) in the Coulomb gauge $\nabla \cdot {\boldsymbol A}=0 \Rightarrow {\boldsymbol A}\cdot {\boldsymbol p}=
{\boldsymbol p}\cdot {\boldsymbol A}$, and
(iii) we choose symmetric gauge, which is quite popular in the many-body theory, defined by
${\boldsymbol A}_0 = B (0, \frac{1}{2} r, 0)$.

\begin{figure}[htbp]
\includegraphics*[width=8cm,height=9cm]{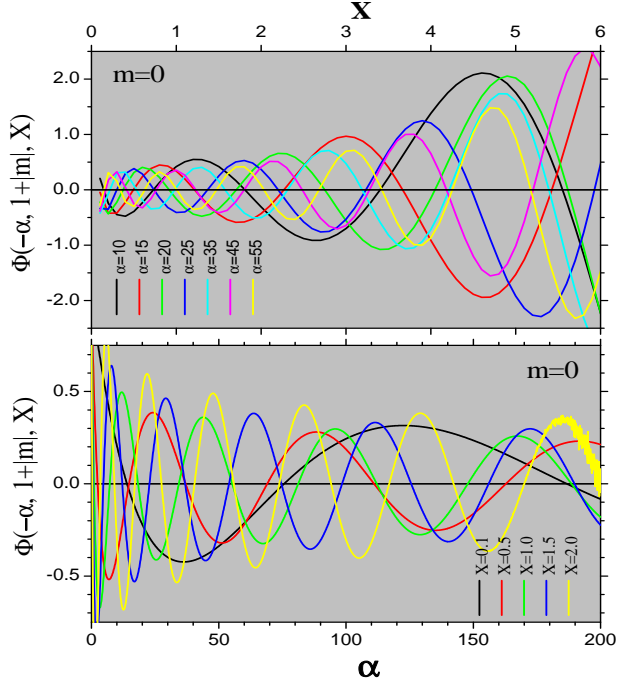}
\caption{(Color online) The plot of the CHF $\Phi (\alpha, 1+|m|, X)$  vs. $X$ (the upper panel) and
$\alpha$ (the lower panel), for $m=0$. [After Kushwaha, Ref. 39].}
\label{fig10}
\end{figure}

Next, we consider, for the sake of simplicity, the quantization energy for the electrons equal to that for the
holes implying that $\omega_e=\omega_o=\omega_h$. Then, in the cylindrical coordinates [($r, \theta, z$)], the Schr\"odinger equation $H_0\psi=\epsilon \psi$ for a quantum wire of radius $R$ and length $L$ is solved to
characterize the system with the eigenfunction $\psi (r,\theta,z)=\psi(r)\,\psi(\theta)\,\psi(z)$, where
\begin{align}
\psi(z)&=\sqrt{\frac{2}{L}}\,\sin(k\,z), \,\,\,\,\,\,\,{\rm with \,\,\,k=n\pi/L}\, ,\\
\psi(\theta)&=\sqrt{\frac{1}{2\pi}}\,\, e^{i\,m\,\theta}\, ,\\
\psi(r)&=\frac{N}{s^{|m|/2}}\,e^{-X/2}\, X^{|m|/2}\,\Phi(-\alpha; 1+|m|; X) ,
\end{align}
where $n$ is an integer, $m=0,\pm 1, \pm 2, ...$ is the azimuthal quantum number, $s=\sqrt{m_+/\mu}$, and $X= r^2/(2\ell^2_c)$, with $\ell_c=\sqrt{\hbar/\mu \Omega_+}$ as the effective magnetic length in the problem,
$m_+=m_e+m_h$ as the total mass, $\mu=m_e m_h/(m_e + m_h)$ as the reduced mass, $\Omega_+=s \Omega=s \sqrt {\omega_{ce}\,\omega_{ch}+4 \omega_o^2}$ as the effective hybrid frequency, and $\omega_{ci}=eB/(m_i c)$ as
the cyclotron frequency. In Eq. (9), $N$ is the normalization coefficient to be determined by the condition
$\int^{R}_{0} dr\, r \mid \psi (...) \mid^2=1$, which yields
\begin{equation}
N^{-2}=\frac{\ell^2_c}{s^{\mid m \mid}}\!\int^{\zeta}_{0}\!\! dX\,e^{-X} X^{\mid m \mid}
                                              \big [\Phi(-\alpha; 1+|m|; X)\big]^2
\end{equation}
where the upper limit $\zeta=R^2/2\ell^2_c$. In Eq. (9), $\Phi(-\alpha; 1+|m|; X)$ is the confluent
hypergeometric function (CHF) [43-45], which is a solution of the Kummer's equation:
$X U^{''}\!+[1+\!\!|m|\!\!-X]\,U'\!+\alpha\,U=0$, where
$\alpha=\frac{1}{\hbar \Omega_+}[\epsilon -\frac{\hbar^2 k^2}{2\mu}-\frac{m}{2}\hbar \Omega_-]-
\frac{1}{2}(1+|m|)$, with
$\Omega_-=(\omega_{ce}-\omega_{ch})$ being the effective cyclotron frequency. The wavefunction $\psi (...)$
obeys the boundary condition $\psi (r=R)=0 \Rightarrow \Phi(-\alpha; 1+|m|; \zeta)=0$, which yields
\begin{equation}
\frac{\epsilon'}{\epsilon_r}=\zeta\,
                  \big [\alpha +\frac{1}{2}\,\big (1+m\,\frac{\Omega_-}{\Omega_+}+\mid m \mid \big)\big]
\end{equation}
where $\epsilon'=\epsilon-\frac{\hbar^2 k^2}{2\mu}$ and $\epsilon_r=2\hbar^2/(\mu R^2)$. It should be pointed
out that the term $\frac{\hbar^2 k^2}{2\mu}$ is only additive and hence inconsequential. What is
important to notice though is that, for a given $m$, $\zeta$ and $\alpha$ are determined self-consistently
where $\Phi(-\alpha; 1+|m|; X)=0$. Figure 10 demonstrates that $\Phi (...)$ is unambiguously a well-behaved
function over a wide range of $X$ and $\alpha$ and that its period is seen to be decreasing with increasing
$X$ or $\alpha$ as the case may be. Since the rest of the illustrative examples require the quantum wire to
be specified, we choose GaAs/Ga$_{1-x}$Al$_x$As system: $m_e=0.067 m_0$, $m_h=0.47 m_0$, and the aspect
ratio $r_a=L/R=10^3$ [46]. Another important parameter in the problem is the reduced magnetic flux
$\phi/\phi_0$: the ratio of the magnetic flux [$\phi=B\pi R^2$] to the
quantum flux [$\phi_0=h c/e$] -- yielding $\phi/\phi_0=\zeta$.

\begin{figure}[htbp]
\includegraphics*[width=8cm,height=9cm]{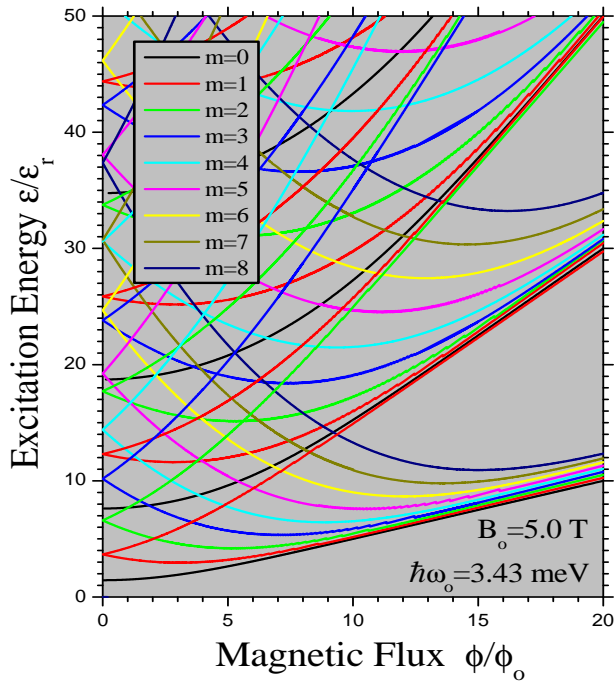}
\centering
\caption{(Color online) The excitation spectrum for a (finite) cylindrical quantum wire subjected to
a magnetic field in the symmetric gauge. The y (x) axis refers to the reduced energy (reduced
magnetic flux). The lower (upper) arm of a wedge corresponds to a negative (positive) value of $m$.
[After Kushwaha, Ref. 39].}
\label{fig11}
\end{figure}

Figure 11 illustrates the excitation spectrum for the cylindrical quantum wire of finite length as a function
of the reduced magnetic flux $\phi/\phi_0$, for several values of $m$. This is a result of (involved)
computation based on Eq. (11) within the strategy as stated above. The parameters used are as listed inside
the picture. For every $m$, the lower (upper) branch of the wedge corresponds to the negative (positive) value
of $m$. It is interesting to see that both arms of each wedge gradually approach the Landau levels at higher
magnetic flux, just as in the case of the Fock-Darwin spectrum of a quantum dot [see, e.g., Ref. 24]. Notice
that this is not a mere coincidence, rather a consequence of the isomorphism in the expression of the
single-particle energies for the two quantum systems.

Next, we proceed to the evaluation of the electrical current density ${\boldsymbol J}$ derived
as [with $m_e < m_h$]
\begin{equation}
{\boldsymbol J}=\frac{i e \hbar}{2 m_e}\,\big (\psi \nabla \psi^{*}-\psi^{*}\nabla \psi \big ) +
\frac{e^2}{m_e c}\,{\boldsymbol A}\,\psi^{*}\psi\, ,
\end{equation}
where the vector potential ${\boldsymbol A}$ is approximated such that
we seek to measure its {\em ac} part (${\boldsymbol A}_1$) just as we do its {\em dc} one (${\boldsymbol A}_0$),
i.e., we express ${\boldsymbol A}_0$ in the symmetric gauge just as before and ${\boldsymbol A}_1 = B(0, 0, r)$.
This is not to say that there aren't other, more subtle, ways to complicate the situation, but since this is
the first paper of its kind, we choose to stick to the bare-bone simplicity -- the complexity will (and should)
come later. Consequently, it is not difficult to split Eq. (12) into its scalar components:
\begin{align}
J_{\theta}&=\Big[\frac{e \hbar m}{m_e r} + \frac{e^2 B r}{2 s m_e} \Big]\,
               \big| \psi (r, \theta, z)\big|^2\nonumber\\
J_{z}&= \frac{e^2 B r}{s m_e}\, \big| \psi (r, \theta, z)\big|^2\, .
\end{align}
The very nature of the magnetic field -- with its charismatic role in localizing the charge carriers in the
plane perpendicular to its orientation -- elucidates why the radial component of the current density $J_{r}=0$.
It is interesting to notice that in the absence of the $z$ component of ${\boldsymbol A}_1$, $J_z$ will be zero
for a quantum wire of {\em finite} length (as is the case here). This is contrary to the case of a wire of an
{\em infinite} length -- where $J_z \ne 0$ even when $[{\boldsymbol A}_1]_z=0$. Equations (13) should, in
principle, provide the clue to the quest in the problem.

\begin{figure}[htbp]
\includegraphics*[width=8cm,height=9cm]{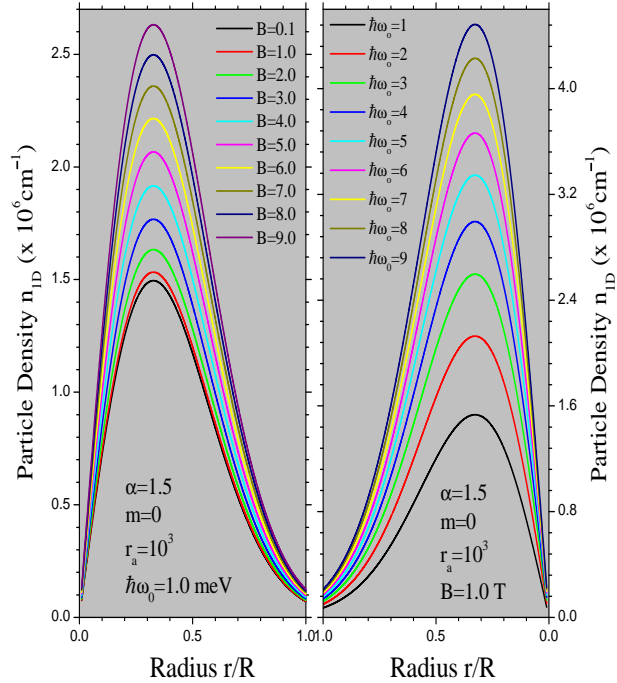}
\centering
\caption{(Color online) The particle density $n_{1D}$ as a function of the reduced radius $r/R$ for the
several values of the magnetic field (left panel) and confinement potential (right panel). The parameters
are as listed inside the picture. Again, it is a GaAs/Ga$_{1-x}$Al$_{x}$As quantum wire.
[After Kushwaha, Ref. 39].}
\label{fig12}
\end{figure}

Figure 12 shows the 1D particle density ($n_{1D}$) in the quantum wire as a function of the reduced radius for
various values of the magnetic field (left panel) and confinement potential (right panel). The confinement
potential in the left panel is $\hbar \omega_0=1.0$ meV and the magnetic field in the right panel is $B=1.0$ T.
The Other parameters are listed inside the picture. The particle density shows a maximum. We were able to find
out that this maximum occurs unequivocally at
$\frac{d}{d r}[n_{1D}]=\frac{d}{d r} [r \Phi^2 (\alpha, 1+|m|, X)]=0\Rightarrow r/R=1/\sqrt{2\,\zeta}=0.3271$.
Since the particle density is fundamental to most of the electronic, optical, and transport properties, we
expect the current features to make a mark in those phenomena, at least, in the cylindrical symmetry. In a
quantum system the particle density has much to do with the Fermi energy in the system. For the fact that
each quantum level can take two electrons with opposite spin, the Fermi energy $\epsilon_F$ of a system of
$\mathcal{N}$ noninteracting electrons for a {\em finite} system at absolute zero temperature is equal to
the energy of the $\frac{1}{2}\mathcal{N}$-th level. As such, one can compute self-consistently the Fermi
energy in a {\em finite} quantum wire containing $\mathcal{N}$ electrons through this expression:
$\mathcal{N}=2 \sum _{m}\theta (\epsilon_F - \epsilon'_{m})$, where $\theta(...)$ is the Heaviside step
function and the summation is only over the azimuthal quantum number $m$. The prime on $\epsilon_{m}$ has the
same meaning as explained before. Because the electronic excitation spectrum happens to be so very intricate
[see Fig. 11], the Fermi energy will not be a smooth function of the magnetic field (or flux). This intricacy
should also lead to complex structures in the magneto-transport phenomena.

\begin{figure}[htbp]
\includegraphics*[width=8cm,height=9.5cm]{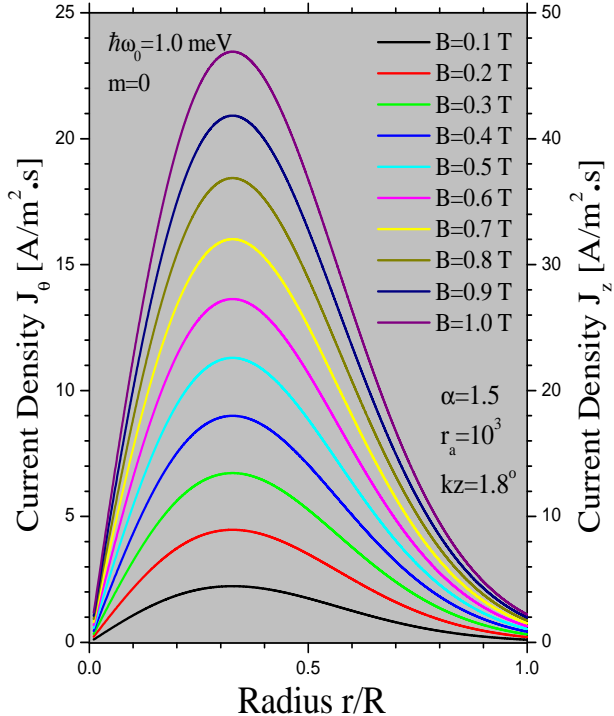}
\centering
\caption{(Color online) The current density $J_{\theta}$ ($J_{z}$) on the left (right) vertical axis
vs. the reduced radius $r/R$ for the several values of the magnetic field. The parameters used are:
$\hbar \omega_0=1.0$ meV, aspect ratio $r_a=1000$, $\alpha=1.5$, and $kz=1.8^o$. For a
GaAs/Ga$_{1-x}$Al$_{x}$As quantum wire just as before. [After Kushwaha, Ref. 39].}
\label{fig13}
\end{figure}

Figure 13 depicts the current density as a function of the reduced radius for several values of the magnetic
field. The parameters used are: the confinement potential $\hbar \omega_0=1.0$ meV, aspect ratio
$r_a=1000$, $\alpha=1.5$, and $kz=1.8^o$. Other material parameters are the same as before. We observe that
the larger the magnetic field, the greater the current density. This is what we should expect intuitively:
the larger the magnetic field, the stronger the confinement of the charge carriers near the axis and hence
greater the current density. The most important aspect this figure reveals is that there is a maximum in the
current density and this maximum is again defined exactly by $r/R=0.3271$. In a certain way, Fig. 12
substantiates the features observed in Fig. 13. This tells us that in a quantum wire the maximum of charge
density lies at $r/R=0.3271$ instead of exactly at the axis. Note that the classical pinch effect in
conventional (3D) SSP [6-19] does not share any such feature. Very close to the axis and to the surface of
the quantum wire, the minimum of the current density is smallest but still nonzero. Traditionally,
$\theta$-pinches ($\Rightarrow J_{\theta}$) tend to resist the plasma instabilities due to the famous
{\em frozen-in-flux} theorem (see Appendix B) [47], whereas $z$-pinches ($\Rightarrow J_z$) tend to favor
the confinement phenomena. The magnitude of $J_{z}$ is double of that of the $J_{\theta}$ just as dictated
by Eqs. (13). We believe that these currents are of moderate strength and would not cause an undue heating of
the two-component plasma in the quantum wires which are generally subjected to experimental observations at
low temperatures.

There are several other interesting and important issues such as the equilibrium, temperature, recombination,
and population inversion the consideration of which would certainly give a better insight into the problem.
The determination of the radial velocity and the pinch radius as a function of time, after the inception
of the pinch, should be significant. The aspect of population inversion enabling the magnetized quantum
wires to act as optical amplifiers has recently been discussed in a different context [48-50]. These issues
are deferred to a future publication.

\begin{figure}[htbp]
\includegraphics*[width=9.75cm,height=7cm]{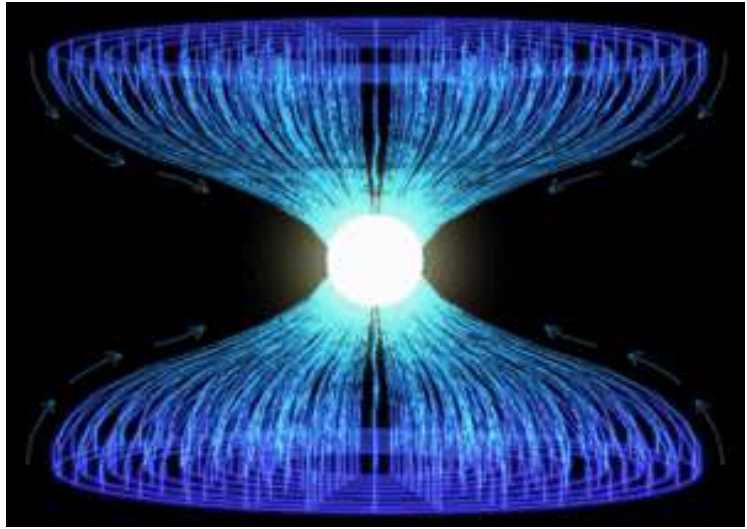}
\centering
\caption{(Color online) Despite the varied views about its shape, size, and life, the radiating sphere of
the sun is known to be the best example of the Birkeland currents focusing into a z-pinched plasma. The
result is a spherically focused and illuminating matter in the center of the solar system: the misunderstood
sun.}
\label{fig14}
\end{figure}

\section{Concluding Remarks}

In summary, we have investigated the quantum analogue of the classical pinch effect in finite quantum wires with
cylindrical symmetry. Since the late 1940s the pinch effect in a gas discharge has been investigated intensively
in laboratories throughout the world, because it offers the possibility of achieving the magnetic confinement of
a hot plasma (a highly ionized gas) necessary for the successful operation of a thermonuclear or fusion reactor.
In solid state plasmas the issues related to confinement, as discussed above, are not encountered. However, since
no system provides an ideal environment in the real world, the SSPs also pose challenges. In a conventional SSP
with no impurities, the thermally excited pair density is a function of temperature only. Any deviation from this (thermal) equilibrium can (and, generally, does) give rise to recombination, which, in turn, affects the pinching
process. The desired maintenance of the equilibrium pair density occurs through various processes such as Coulomb interactions and particle-lattice interactions, which are not independent and hence cause unintended consequences.
The quantum wires -- just as other low-dimensional structures such as quantum wells and quantum dots -- are the
systems in which most of the experiments are performed at low (close to zero) temperatures. Therefore, the risks
of thermal non-equilibrium and recombination are much smaller than those ordinarily encountered in conventional
SSP. This implies that two-component quantum wires at low temperatures offer an ideal platform for the realization
of the quantum pinch effects.

\begin{figure}[htbp]
\includegraphics*[width=9.5cm,height=7cm]{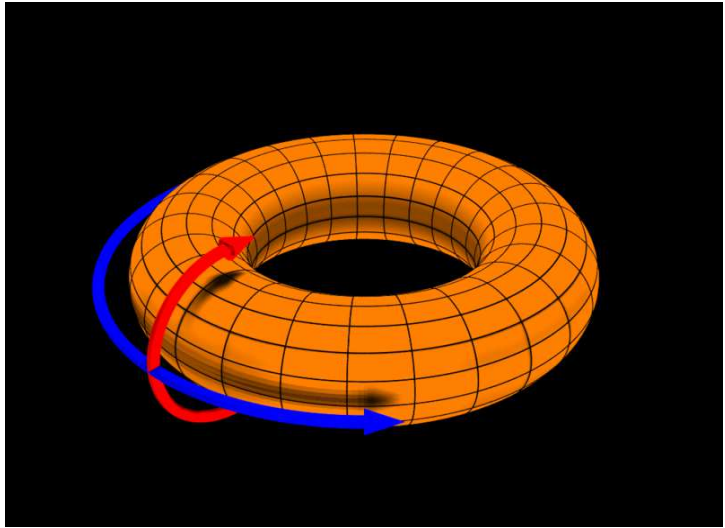}
\centering
\caption{(Color online) A toroidal tube depicting the poloidal ($\theta$) direction [red arrow]
and the toroidal ($\phi$) direction [blue arrow]. Notably absent is the radial coordinate which
starts from the center of the tube and points out. Nothing is perfect: this torus breaks the
$\theta$ symmetry. [After Dave Burke (2006)].}
\label{fig15}
\end{figure}

Myriad of applications of quantum pinches bud out from the very thought of the self-focused, two-component plasma
in a quantum wire with cylindrical symmetry. The plasma by definition is electrically conductive implying that it
responds strongly to electromagnetic fields. The self-focusing (or pinching) only adds to the response. The
quantum wire brings all this to the nanoscale. Therefore, we are designing an electronic device that can (and
will) control the particle beams at the nanoscale. Potential applications include extremely refined nanoswitches,  nanoantennas, optical amplifiers, and precise particle-beam nanoweapons, just to name a few. The greatest
advantage of the quantum pinch effect over its classical counterpart is that it offers a Gaussian-like cycle of
operation with two minima passing through a maximum. The smooth functionality of the plasma devices is, however,
based on a single tenet: there must be means not only to produce it, but also to sustain it.

According to one retelling, Prometheus -- the (mythological) Greek God -- gave mankind not just the fire but the
very arts of civilization itself –- from writing to medicine to astronomy. Fusion is the Promethean fire that
has lit the heavens for billions of years. It is the ultimate clean, renewable energy resource, burning seawater
as fuel, and producing no carbon dioxide, smoke, or radioactive waste. What it does produce is a prodigious and
inexhaustible source of energy –- in principle, a teaspoon of hydrogen can generate as much energy as 10 metric
tons of coal. Fusion -– it powers the Sun; it powers the Stars; it powers the Galaxy. Why can we not make it
power the Earth as well?


\begin{acknowledgments}
The author feels grateful to Loren N. Pfeiffer who affirmed the feasibility of achieving the aspect
ratio of $r_a=1000$ in the semiconducting quantum wires within the current technology and for the
fruitful discussions. He would also like to express his sincere thanks to Alex Maradudin, Allan
MacDonald, Bahram Djafari, H. Sakaki, Bob Camley, Jun Kono, Naomi Halas, Peter Nordlander, and Chizuko
Dutta for useful communications and stimulating discussions on various aspects related to the subject.
He is appreciative of Kevin Singh for the invaluable help with the software during the course of this
investigation. Finally, he should like to thank Dr. K.K. Phua, Editor-in-Chief, for the invitation to
contribute to this subject, which is still in its infancy, and for his great patience.
\end{acknowledgments}
\newpage
\appendix

\section{The Equilibrium Analysis of The Pinch Geometries}

\subsection{One Dimension}

One dimensional geometry is generally a cylindrical tube which is symmetrical in the axial ($z$) direction
as well as in the azimuthal ($\theta$) direction. Three pinch geometries generally studied in one dimension
are: the $\theta$-pinch, the Z-pinch, and the screw pinch. The one-dimensional pinches are known after the
direction in which the current flows. Note that we assume the physical geometry to be made up of
non-magnetic materials, which implies ${\bf B}\equiv {\bf H}$ in the Maxwell equations. Also, we restrict
ourselves to the situation shielded by $\nabla\cdot{\bf J}=0$, i.e., where we discard the displacement
current in the fourth Maxwell equation. In addition, the magnetic field ${\bf B}$ is only the function of
$r$, i.e., ${\bf B}=B_z (r) \hat{z} + B_{\theta}(r) \hat{\theta}$.

\subsubsection{The $\theta$-pinch}

The $\theta$-pinch has a magnetic field oriented in the z direction [i.e., ${\bf B} = B_z (r)\,\hat{z}$].
Then employing Ampere's law yields
\begin{align}
\frac{1}{c}\,{\bf J}
= \frac{1}{4\pi}\,(\nabla \times {\bf B})
= \frac{1}{4\pi r}\,\frac{d}{d \theta}B_z(r)\, \hat{r} - \frac{1}{4\pi}\,\frac{d}{d r}B_z(r)\,\hat{\theta}
\end{align}
Knowing that ${\bf B}$ is only a function of $r$, this simplifies to
\begin{align}
\frac{1}{c}\,{\bf J}
= - \frac{1}{4\pi}\,\frac{d}{d r}B_z(r)\,\hat{\theta}
\end{align}
So, ${\bf J}$ points in the $\theta$ direction. Thus, the equilibrium condition [$\nabla p=\frac{1}{c}\,$
(${\bf J}\times{\bf B}$)] reads:
\begin{align}
\frac{d}{d r}\Big(p + \frac{1}{8\pi}\,B^2_z\Big)=0
\end{align}
Symbol $p$ refers to the kinetic pressure of the plasma. Notice that the $\theta$-pinch tends to resist the
plasma instabilities. This is due in part to the Alfv\'en's theorem (or frozen-in-flux theorem)
[see Appendix B].

\subsubsection{The $Z$-pinch}

The $Z$-pinch has a magnetic field oriented in the $\theta$ direction
[i.e., ${\bf B} = B_{\theta} (r)\hat{\theta}$]. Again, employing Ampere's law yields
\begin{align}
\frac{1}{c}\,{\bf J}
= \frac{1}{4\pi}\,(\nabla \times {\bf B})
= \frac{1}{4\pi r}\,\frac{d}{d r}(r B_{\theta} (r))\,\hat{z} - \frac{1}{4\pi}\,\frac{d}{d z}B_{\theta}(r)\,
                              \hat{r}
\end{align}
Since ${\bf B}$ is only a function of $r$, this reduces to
\begin{align}
\frac{1}{c}\,{\bf J}
= \frac{1}{4\pi r}\,\frac{d}{d r}(r B_{\theta}(r))\,\hat{z}
\end{align}
So, ${\bf J}$ points in the z direction. Thus, the equilibrium condition [$\nabla p=\frac{1}{c}\,$
(${\bf J}\times{\bf B}$)] reads:
\begin{align}
\frac{d}{d r}\Big(p + \frac{1}{8\pi}\,B^2_{\theta}\Big) + \frac{1}{4\pi r}\,B^2_{\theta}=0
\end{align}
Since particles in the plasma basically follow the magnetic field lines, the Z-pinch leads them around in
circles. Therefore, the Z-pinch offers excellent confinement characteristics [see Fig. 14].

\subsubsection{The screw pinch}

In the screw pinch, the magnetic field has a $\theta$ component as well as a z component
[i.e., ${\bf B}=B_z (r) \hat{z} + B_{\theta}(r) \hat{\theta}$]. The screw pinch is explicitly an effort to
merge the stability aspects of the $\theta$-pinch and the confinement aspects of the Z-pinch. Referring once
again to Ampere's law
\begin{align}
\frac{1}{c}\,{\bf J}
= \frac{1}{4\pi}\,(\nabla \times {\bf B})
= \frac{1}{4\pi r}\,\frac{d}{d r}(r B_{\theta}(r))\,\hat{z}- \frac{1}{4\pi}\,\frac{d}{d r}B_z(r)\,\hat{\theta}
\end{align}
So, this time, ${\bf J}$ has a component in the z direction and a component in the $\theta$ direction. Thus,
the equilibrium condition [$\nabla p=\frac{1}{c}\,$(${\bf J}\times{\bf B}$)] reads:
\begin{align}
\frac{d}{d r}\Big(p + \frac{1}{8\pi}\,\big(B^2_z+B^2_{\theta}\big)\Big) + \frac{1}{4\pi r}\,B^2_{\theta}=0
\end{align}
It is noteworthy that with the neglect of the displacement current in the fourth Maxwell equation, it is quite
appropriate to ignore Coulomb's law as well. This is because the electric field is completely determined by
the curl equations and Ohm's law [${\bf J}=\sigma\,{\bf E}$]. If the displacement current is retained and
$\nabla\cdot{\bf E}=4\pi \rho_e$ is taken into account, corrections of only the order of $v^2/c^2$ result. For
the normal MHD problems these are completely negligible. That is to say that the whole analysis in this Appendix
lies within the legitimate realm of MHD.

\subsection{Two Dimensions}

A common problem with one-dimensional pinches is the end losses. Most of the motion of particles is along the
magnetic field. With the $\theta$-pinch and the screw-pinch, this leads particles out of the end of the machine
very quickly, resulting in a loss of mass and energy. Besides, the Z-pinch has severe instability problems.
Though particles can be reflected to some extent with magnetic mirrors, even these allow considerable number of
them to still pass through. A common remedy of controlling these end losses is to bend the cylinder around into
a torus [see Fig. 15]. Unfortunately, this breaks $\theta$ symmetry, as paths on the inner portion of the torus
are shorter than the those on the outer portion. Thus a new theoretical framework is needed. This gives rise to
the famous Grad-Shafranov equation numerical solutions of which have yielded some equilibria, most notably that
of the reversed pinch.

\section{Alfv\'en's Frozen-in-Flux Theorem}

For the plasma assumed to be a perfect conductor [with the conductivity $\sigma\rightarrow \infty$], the Ohm's
law renders:
\begin{align}
E + \frac{1}{c}\,({\bf V}\times {\bf B})=0
\end{align}
This is sometimes referred to as the flux freezing equation. The nomenclature comes about because Eq. (B1)
implies that the magnetic flux through any closed contour in the plasma, each elemnt of which moves with the
local velocity, is a conserved quantity. In order to verify this assertion, let us consider the magnetic flux
$\phi_m$, through a contour $C$, which is co-moving  with the plasma:
\begin{align}
\phi_m=\int_S {\bf B}\cdot d{\bf S}\, ,
\end{align}
where ${\bf S}$ is some surface which spans ${\bf C}$. The time rate of change of $\phi_m$ is made up of two
parts. First, there is the part due to the time variation of ${\bf B}$ over the surface ${\bf S}$. This can be
written as
\begin{align}
\frac{\partial \phi_m}{\partial t}\Big|_1 =\int_{\bf S} \frac{\partial {\bf B}}{\partial t}\cdot d{\bf S}
\end{align}
Using the third curl Maxwell equation, this becomes
\begin{align}
\frac{\partial \phi_m}{\partial t}\Big|_1 =-c\,\int_{\bf S} (\nabla\times{\bf E})\cdot d{\bf S}
\end{align}
Second, there is the part due to the motion of ${\bf C}$. If $d{\bf l}$ is an element of ${\bf C}$, then
$V \times d{\bf l}$ is the area swept out by $d{\bf l}$ per unit time. Hence the flux crossing this area is
${\bf B}\cdot$(${\bf V}\times d{\bf l}$). It follows that
\begin{align}
\frac{\partial \phi_m}{\partial t}\Big|_2
=\int_C {\bf B}\cdot({\bf V}\times d{\bf l})=\int_{\bf C} ({\bf B}\times{\bf V})\cdot d{\bf l}
\end{align}
Using Stokes' theorem, this becomes
\begin{align}
\frac{\partial \phi_m}{\partial t}\Big|_2 =-\int_{\bf S} \nabla\times ({\bf V}\times{\bf B})\cdot d{\bf S}
\end{align}
Hence, the total time rate of change of $\phi_m$ is given by, from Eqs. (B4)  and (B6),
\begin{align}
\frac{d \phi_m}{d t} =-c\,\int_{\bf S}
                             \nabla\times \big[{\bf E}+\frac{1}{c}({\bf V}\times{\bf B})\big]\cdot d{\bf S}
\end{align}
Given the condition in Eq. (B1), Eq. (B7) is led to conclude that
\begin{align}
\frac{d \phi_m}{d t} = 0
\end{align}
In other words, $\phi_m$ remains constant in time for an arbitrary contour. This, in turn, implies that the
magnetic field-lines must move with the plasma. That is to say that the filed-lines are frozen into the
plasma. Since the magnetic filed-lines can be regarded as infinitely thin flux-tubes, we conclude that MHD
plasma motion also maintains the integrity of field-lines. That means magnetic field-lines embedded in an
MHD plasma can never break and reconnect: i.e., MHD forbids any change in the topology of the field-lines.
It turns out that this is an extremely restrictive constraint.

\newpage

\end{document}